\documentclass[fleqn,usenatbib]{mnras}
\usepackage{newtxtext,newtxmath}
\usepackage[T1]{fontenc}
\usepackage{ae,aecompl}

\usepackage{graphicx}	
\usepackage{amsmath}	
\usepackage{amssymb}	

\newcommand{\rmaxs}{\ifmmode{R_{\rm{\sigma}}^{\rm{max}}}\else{$R_{\rm{\sigma}}^{\rm{max}}$}\fi}
\newcommand{\recirc}{\ifmmode{R_{\rm{e,c}}}\else{$R_{\rm{e,c}}$}\fi}
\newcommand{\re}{\ifmmode{R_{\rm{e}}}\else{$R_{\rm{e}}$}\fi}
\newcommand{\ret}{\ifmmode{R_{\rm{e/2}}}\else{$R_{\rm{e/2}}$}\fi}
\newcommand{\retwo}{\ifmmode{R_{\rm{e/2}}}\else{$2R_{\rm{e}}$}\fi}
\newcommand{\aee}{\ifmmode{a_{\rm{e}}}\else{$a_{\rm{e}}$}\fi}
\newcommand{\kpa}{\ifmmode{PA_{\rm{kin}}}\else{$PA_{\rm{kin}}$}\fi}
\newcommand{\ee}{\ifmmode{\epsilon_{\rm{e}}}\else{$\epsilon_{\rm{e}}$}\fi}

\newcommand{\lr}{\ifmmode{\lambda_R}\else{$\lambda_{R}$}\fi}
\newcommand{\lre}{\ifmmode{\lambda_{R_{\rm{e}}}}\else{$\lambda_{R_{\rm{e}}}$}\fi}
\newcommand{\lret}{\ifmmode{\lambda_{R_{\rm{e/2}}}}\else{$\lambda_{R_{\rm{e/2}}}$}\fi}
\newcommand{\lretwo}{\ifmmode{\lambda_{2R_{\rm{e}}}}\else{$\lambda_{2R_{\rm{e}}}$}\fi}

\newcommand{\flr}{\ifmmode{f_{\lambda_{R}}}\else{$f_{\lambda_{R_{\rm{e}}}}$}\fi}
\newcommand{\flre}{\ifmmode{f_{\lambda_{R_{\rm{e}}}}}\else{$f_{\lambda_{R_{\rm{e}}}}$}\fi}
\newcommand{\flret}{\ifmmode{f_{\lambda_{R_{\rm{e/2}}}}}\else{$f_{\lambda_{R_{\rm{e/2}}}}$}\fi}

\newcommand{\vs}{\ifmmode{V / \sigma}\else{$V / \sigma$}\fi}
\newcommand{\vse}{\ifmmode{(V / \sigma)_{\rm{e}}}\else{$(V / \sigma)_{\rm{e}}$}\fi}
\newcommand{\vset}{\ifmmode{(V / \sigma)_{\rm{e/2}}}\else{$(V / \sigma)_{\rm{e/2}}$}\fi}
\newcommand{\vsetwo}{\ifmmode{(V / \sigma)_{\rm{2e}}}\else{$(V / \sigma)_{\rm{2e}}$}\fi}

\newcommand{\vobs}{\ifmmode{V_{\rm{obs}}}\else{$V_{\rm{obs}}$}\fi}
\newcommand{\sobs}{\ifmmode{\sigma_{\rm{obs}}}\else{$\sigma_{\rm{obs}}$}\fi}

\newcommand{\kms}{\ifmmode{\,\rm{km}\, \rm{s}^{-1}}\else{$\,$km$\,$s$^{-1}$}\fi}
\newcommand{\msun}{\ifmmode{M_{\odot}}\else{$M_{\odot}$}\fi}

\newcommand{\at}{\ifmmode{\rm{ATLAS}^{\rm{3D}}}\else{ATLAS$^{\rm{3D}}$}\fi}

\title[Revising the Fraction of Slow Rotators in IFS Galaxy Surveys]{The SAMI Galaxy Survey: Revising the Fraction of Slow Rotators in IFS Galaxy Surveys}

\author[Jesse van de Sande]{
Jesse van de Sande$^{1},$
Joss Bland-Hawthorn$^{1},$
Sarah Brough$^{2,3},$
Scott M. Croom$^{1,3},$
\newauthor
Luca Cortese$^{4},$
Caroline Foster$^{5},$
Nicholas Scott$^{1,3},$ 
Julia J. Bryant$^{1,3,5},$
\newauthor
Francesco d'Eugenio$^{6},$
Chiara Tonini$^{7},$
Michael Goodwin$^{5},$
Iraklis S. Konstantopoulos$^{5,8},$
\newauthor
Jon S. Lawrence$^{5},$
Anne M. Medling$^{6,9,10},$
Matt S. Owers$^{5,11},$ 
Samuel N. Richards$^{12},$
\newauthor
Adam L. Schaefer$^{1,3,5},$
Sukyoung K. Yi$^{13}$
\\
$^{1}${Sydney Institute for Astronomy, School of Physics, A28, The University of Sydney, NSW, 2006, Australia}\\
$^{2}${School of Physics, University of New South Wales, NSW 2052, Australia}\\
$^{3}${ARC Centre of Excellence for All-Sky Astrophysics (CAASTRO)}\\
$^{4}${International Centre for Radio Astronomy Research, The University of Western Australia, 35 Stirling Highway, Crawley WA 6009, Australia}\\
$^{5}${Australian Astronomical Observatory, PO Box 915, North Ryde NSW 1670, Australia}\\
$^{6}${Research School of Astronomy and Astrophysics, Australian National University, Canberra ACT 2611, Australia}\\
$^{7}${School of Physics, the University of Melbourne, Parkville, VIC 3010, Australia}\\
$^{8}${Atlassian 341 George St Sydney, NSW 2000}\\
$^{9}${Cahill Center for Astronomy and Astrophysics California Institute of Technology, MS 249-17 Pasadena, CA 91125, USA} \\
$^{10}${Hubble Fellow}\\
$^{11}${Department of Physics and Astronomy, Macquarie University, NSW 2109, Australia}\\
$^{12}${SOFIA Operations Center, USRA, NASA Armstrong Flight Research Center, 2825 East Avenue P, Palmdale, CA 93550, USA}\\
$^{13}${Department of Astronomy and Yonsei University Observatory, Yonsei University, Seoul 120-749, Republic of Korea}\\
}
\date{}

\pubyear{2017}

\begin{document}
\label{firstpage}
\pagerange{\pageref{firstpage}--\pageref{lastpage}}
\maketitle

\begin{abstract}
The fraction of galaxies supported by internal rotation compared to galaxies stabilized by internal pressure provides a strong constraint on galaxy formation models. In integral field spectroscopy surveys, this fraction is biased because survey instruments typically only trace the inner parts of the most massive galaxies. We present aperture corrections for the two most widely used stellar kinematic quantities $V/\sigma$ and $\lambda_{R}$ (spin parameter proxy). Our demonstration involves integral field data from the SAMI Galaxy Survey and the ATLAS$^{\rm{3D}}$ Survey. We find a tight relation for both $V/\sigma$ and $\lambda_{R}$ when measured in different apertures that can be used  as a linear transformation as a function of radius, i.e., a first-order aperture correction. In degraded seeing, however, the aperture corrections are more significant as the steeper inner profile is more strongly affected by the point spread function than the outskirts. We find that $V/\sigma$ and $\lambda_{R}$ radial growth curves are well approximated by second order polynomials. By only fitting the inner profile (0.5$R_{\rm{e}}$), we successfully recover the profile out to one $R_{\rm{e}}$ if a constraint between the linear and quadratic parameter in the fit is applied. However, the aperture corrections for $V/\sigma$ and $\lambda_{R}$ derived by extrapolating the profiles perform as well as applying a first-order correction. With our aperture-corrected $\lambda_{R}$ measurements, we find that the fraction of slow rotating galaxies increases with stellar mass. For galaxies with $\log M_{*}$/\msun\ $>$ 11, the fraction of slow rotators is $35.9\pm4.3$ percent, but is underestimated if galaxies without coverage beyond one $R_{\rm{e}}$ are not included in the sample ($24.2\pm5.3$ percent). With measurements out to the largest aperture radius the slow rotator fraction is similar as compared to using aperture corrected values ($38.3\pm4.4$ percent). Thus, aperture effects can significantly bias stellar kinematic IFS studies, but this bias can now be removed with the method outlined here.
\end{abstract}





\section{Introduction}

Dynamical studies of stars in galaxies are key to understanding their individual formation history \citep[e.g., ][]{dezeeuw1991,cappellari2016}. Stellar absorption line spectroscopy revealed for the first time that certain galaxies are rotating \citep{slipher1914,pease1916}, well before the discovery was made in our own Galaxy \citep{oort1927}. Later studies confirmed that most disc galaxies show rotation \citep[see e.g., ][]{vanderkruit1978}. Kinematic observations using long-slit spectroscopy of elliptical galaxies discovered that: luminous ellipticals rotate slowly \citep{illingworth1977,bertola1989,binney1978}, bulges of spiral galaxies show rapid rotation \citep{fillmore1986,illingworth1982,kormendy1982,mcelroy1983,whitmore1984}, and that intrinsically faint ellipticals rotate as rapidly as bulges \citep{davies1983}.

The introduction of the SAURON integral field spectrograph \citep[IFS][]{bacon2001}, and the subsequent SAURON \citep{dezeeuw2002} and \at\ survey \citep{cappellari2011a}, led to a more quantified classification of rotation by using two-dimensional (2D) measurements of \vs, the flux weighted ratio between the projected velocity and velocity dispersion, and the spin parameter proxy \lr\ \citep{emsellem2007,cappellari2007}. Galaxies with $\lre>0.31\sqrt{\epsilon}$ were classified as fast rotators; galaxies below this limit as slow rotators \citep{emsellem2011}. The fast/slow rotator separation was motivated by a classification based on kinematic features of the velocity field \citep{krajnovic2011}. The \at\ results indicate that galaxies with regular rotation fields are almost always fast rotators, but non-regular rotators either showed no indication of rotation, revealed signs of kinematically decoupled cores, or counter-rotating discs.

Kinematic classifications of galaxies, however, are sensitive to the aperture in which \vs\ or \lr\ are measured. Due to the limited angular size of integral field spectrographs (IFS), almost all surveys have observed a fraction of galaxies where the aperture does \emph{not} extend to one effective radius (\re): 57 percent of galaxies in the \at\ Survey \citep{emsellem2011}, ~10 percent in the CALIFA survey \citep{falconbarroso2017}, and 24 percent in the SAMI Galaxy Survey \citep{vandesande2017}. As \lr\ growth curves are typically steeply increasing within one \re\ \citep{emsellem2011,fogarty2014,foster2016,veale2017,vandesande2017}, with the exception of kinematically decoupled cores or counter-rotating discs \citep{emsellem2011}, this implies that the fast/slow classification becomes more uncertain if a mix of projected apertures are used\footnote{For example, in \citet{emsellem2011}, \lre\ is quoted and used regardless of the \re\ coverage factor}.

One solution is to implement different selection criteria depending on the aperture \citep[e.g., ][]{fogarty2014,brough2017}. Another solution is to aperture correct \vs\ and \lr\ to one \re, similar to methods applied to velocity dispersions, where aperture corrections have been measured and applied to low and high-redshift galaxies for more than two decades \citep{jorgensen1995,cappellari2006,vandesande2013,falconbarroso2017}.

In order to estimate aperture corrections for \vs\ and \lr\ we require 2D stellar kinematic measurements with sufficient sampling within \re. A wealth of such stellar kinematic data is
becoming available from large multi-object IFS surveys such as the SAMI Galaxy Survey \citep[Sydney-AAO Multi-object Integral field spectrograph; N $\sim3600$; ][]{croom2012,bryant2015} and the SDSS-IV MaNGA Survey \citep[Sloan Digital Sky Survey Data; Mapping Nearby Galaxies at APO; N $\sim10000$; ][]{bundy2015}, and other single-shot IFS surveys, such as the \at\ survey \citep[N=260; ][]{cappellari2011a}, the CALIFA Survey \citep[N $\sim480-600$; ][]{sanchez2012}, and the MASSIVE Survey \citep[N$\sim100$; ][]{ma2014,veale2017}.  
Given the large spread in aperture size between these IFS surveys, a simple aperture correction method is urgently required to spatially homogenise all samples.

In this paper, we present stellar kinematic aperture corrections for \vs\ and \lr\ from the SAMI Galaxy Survey and the publicly available data from the \at\ Survey. The paper is organized as follows: Section \ref{sec:data} describes the SAMI Galaxy Survey and \at\ data, and our method for extracting the stellar kinematics. In Section \ref{sec:ac}, we explore aperture corrections using a simple method (Section \ref{subsec:simple_ac}) and using growth curves (Section \ref{subsec:curve_growth_ac}). With the SAMI and \at\ aperture-corrected data we study the fraction of fast and slow rotators in Section \ref{sec:application}, and summarize and conclude in Section \ref{sec:conclusions}. Throughout the paper we assume a $\Lambda$CDM cosmology with $\Omega_\mathrm{m}$=0.3, $\Omega_{\Lambda}=0.7$, and $H_{0}=70$ km s$^{-1}$ Mpc$^{-1}$.



\section{Data}
\label{sec:data}

\subsection{SAMI Galaxy Survey}
\subsubsection{Observations and Target Selection}
SAMI is a multi-object IFS mounted at the prime focus of the 3.9m Anglo Australian Telescope (AAT). It employs 13 of the revolutionary imaging fibre bundles, or \textit{hexabundles} \citep{blandhawthorn2011,bryant2011,bryant2012a,bryant2014}, which are made out of 61 individual fibres with 1\farcs6 angle on sky. Each hexabundle covers a $\sim15^{\prime\prime}$ diameter region on the sky, has a maximal filling factor of 75\%, and is deployable over a 1$^\circ$ diameter field of view. All 819 fibres, including 26 individual sky fibres, are fed into the AAOmega dual-beamed spectrograph \citep{saunders2004, smith2004, sharp2006}. 

The SAMI Galaxy Survey \citep{croom2012, bryant2015} aims to observe 3600 galaxies, covering a broad range in galaxy stellar mass (M$_* = 10^{8}-10^{12}$\msun) and galaxy environment (field, groups, and clusters). The redshift range of the survey, $0.004<z<0.095$, results in spatial resolutions of 1.6 kpc per fibre at $z=0.05$. Field and group targets were selected from four volume-limited galaxy samples derived from cuts in stellar mass in the Galaxy and Mass Assembly (GAMA) G09, G12 and G15 regions \citep{driver2011}. GAMA is a major campaign that combines a large spectroscopic survey of $\sim$300,000 galaxies carried out using the AAOmega multi-object spectrograph on the AAT, with a large multi-wavelength photometric data set. Cluster targets were obtained from eight high-density cluster regions sampled within radius $R_{200}$ with the same stellar mass limit as for the GAMA fields \citep{owers2017}. 

For the SAMI Galaxy Survey, the 580V and 1000R grating are used in the blue (3750-5750\AA) and red (6300-7400\AA) arm of the spectrograph, respectively. This results in a resolution of R$_{\rm{blue}}\sim 1810$ at 4800\AA, and R$_{\rm{red}}\sim4260$ at 6850\AA\ \citep{vandesande2017}. In order to create data cubes with 0\farcs5 spaxel size, all observations are carried out using a six to seven position dither pattern \citep{sharp2015,allen2015}.

\subsubsection{Ancillary Data}

For galaxies in the GAMA fields, we use the aperture matched $g$ and $i$ photometry from the GAMA catalog  \citep{hill2011,liske2015}, measured from reprocessed SDSS Data Release Seven \citep{york2000, kelvin2012}, to derive $g-i$ colours. For the cluster environment, photometry from the SDSS \citep{york2000} and VLT Survey Telescope ATLAS imaging data are used \citep{shanks2013,owers2017}. 

Effective radii, ellipticities, and positions angles are derived using the Multi-Gaussian Expansion \citep[MGE;][]{emsellem1994,cappellari2002} technique and the code from \citet{scott2009} on imaging from the GAMA-SDSS \citep{driver2011}, SDSS \citep{york2000}, and VST \citep{shanks2013,owers2017}. 
We define \re\ as the semi-major axis effective radius, and the ellipticity of the galaxy within one effective radius as $\epsilon_{\rm{e}}$, measured from the best-fitting MGE model. For more details, we refer to D'Eugenio et al. (in prep).

\subsubsection{Stellar Kinematics}
\label{subsec:stelkin_sami}

Stellar kinematics are measured from the SAMI data by using the penalized pixel fitting code \citep[pPXF;][]{cappellari2004} as described in \citet{vandesande2017}. All 1380 unique galaxy cubes, i.e., not including repeat observations, that make up the SAMI Galaxy Survey internal v0.9.1 data release (from observations up to December 2015) are fit with the SAMI stellar kinematic pipeline, assuming a Gaussian line of sight velocity distribution (LOSVD), i.e., extracting only the stellar velocity and stellar velocity dispersion.

In summary, we first convolve the red spectra to match the instrumental resolution in the blue. The blue and red spectra are then rebinned onto a logarithmic wavelength scale with constant velocity spacing (57.9 \kms), using the code \textsc{log\_rebin} provided with the \textsc{pPXF} package. We use annular binned spectra for deriving local optimal templates from the MILES stellar library \citep{sanchezblazquez2006}, which consists of 985 stars spanning a large range in stellar atmospheric parameters.

After the optimal template is constructed for each annular bin, we run \textsc{pPXF} three times on each galaxy spaxel. One time for getting precise measure of the noise scaling from the residual of the fit, a second time for the masking of emission lines and clipping outliers using the CLEAN parameter in \textsc{pPXF}, and a third time to extract the velocity and velocity dispersion. In the third iteration \textsc{pPXF} is allowed to use the optimal templates from the annular bin in which the spaxel is located, as well as the optimal templates from neighbouring annular bins. We use a 12th order additive Legendre polynomial to remove residuals from small errors in the flux calibration. Finally, the uncertainties on the LOSVD parameters are estimated from 150 simulated spectra.

As demonstrated in \citet{vandesande2017}, for the SAMI Galaxy Survey we impose the following quality criteria to the stellar kinematic data: signal-to-noise (S/N) $>3$\AA$^{-1}$, \sobs $>$ FWHM$_{\rm{instr}}/2 \sim 35$\kms\ where the FWHM is the full-width at half-maximum, $V_{\rm{error}}<30$\kms, and $\sigma_{\rm{error}} < \sobs *0.1 + 25$\kms\ \citep[Q$_1$ and Q$_2$ from][]{vandesande2017}. From a visual inspection of all 1380 SAMI kinematic maps we flag and exclude 41 galaxies with irregular kinematic maps due to nearby objects or mergers that influence the stellar kinematics of the main object. We furthermore exclude 369 galaxies where $\re<1\farcs5$ or where either \re\ or the radius out to which we can accurately measure the stellar kinematics is less than the half-width at half-maximum of the PSF (HWHM$_{\rm{PSF}}$). This brings the final number of galaxies with usable stellar velocity and stellar velocity dispersion maps to 970.

\subsection{\at\ Survey}

The SAURON survey \citep{dezeeuw2002} and \at\ Survey \citep{cappellari2011a} have a complete combined sample of 260 early-type galaxies within the local (42 Mpc) volume observed with the SAURON spectrograph \citep{bacon2001}. For the SAURON survey a spectral resolution of 4.2\AA\ FWHM ($\sigma_{\rm{instr}}$ = 105 \kms) was adopted, covering the wavelength range 4800-5380\AA, whereas for the \at\ survey, galaxies were observed with a higher resolution of 3.9\AA\ FWHM ($\sigma_{\rm{instr}}$ = 98\kms). With a pixel scale of 0\farcs7, the average spatial resolution is $\sim$0.1kpc at $\sim$20Mpc. The data were Voronoi binned \citep{cappellari2003} with a target signal-to-noise of 40. As described in \citet{cappellari2011a}, the stellar kinematics were extracted using \textsc{pPXF} with stellar templates from the MILES stellar library \citep{sanchezblazquez2006}. 

Here, we use the \at\ Survey's publicly available online data\footnote{http://www-astro.physics.ox.ac.uk/atlas3d/}. In particular, we use the unbinned data cubes (V1.0) and the 2D Voronoi binned stellar kinematic maps \citep{emsellem2004,cappellari2011a}\footnote{Unbinned stellar kinematic measurements are not available.}. Galaxy NGC 0936 is excluded from the sample as no unbinned data is available. We adopt the circularised size measurements from \citet{cappellari2011a}, which are corrected to semi-major axis effective radii using the global ellipticities from \cite{krajnovic2011} ($\re=\recirc/\sqrt{1-\epsilon}$). Ellipticities at one effective radius are from \citet{emsellem2011}, and position angles from \cite{krajnovic2011}. Furthermore, we calculate stellar masses from the $R$ band luminosity and mass-to-light ratio as presented in \citet{cappellari2013a,cappellari2013b} and correct these to a \citet{chabrier2003} IMF.



\section{Aperture Corrections for \vs\ \& \lr}
\label{sec:ac}

In this section, we first discuss why aperture corrections are needed by showing the largest stellar kinematic aperture radius as a function of stellar mass. Next, we explore two different approaches for calculating aperture corrections: corrections from a simple relation between \vs, or \lr\ at different radii, and corrections extrapolated from radial growth curves.

\subsection{Largest Aperture Radius}
\label{subsec:maxaperre}
For each galaxy, we calculate the largest aperture radius out to which the stellar kinematic data meet our quality criteria. This \rmaxs\ is defined as the semi-major axis of an ellipse where at least 85\% of the spaxels meet our quality control criteria. The axis ratio and position angle of the ellipse are obtained from the 2D MGE fits to the imaging data. For the SAMI Galaxy Survey data, we use the unbinned velocity and velocity dispersion maps as described in Section \ref{subsec:stelkin_sami}. For the \at\ data, the unbinned flux maps are combined with the Voronoi binned stellar kinematic data. To translate the Voronoi binned stellar kinematic data back to the unbinned grid, we assign the same velocity and velocity dispersion of a Voronoi bin to all spaxels within the same Voronoi bin. All spaxels that are flagged or impacted by cosmic rays, stars, or nearby objects are masked.

\begin{figure}
	\includegraphics[width=\linewidth]{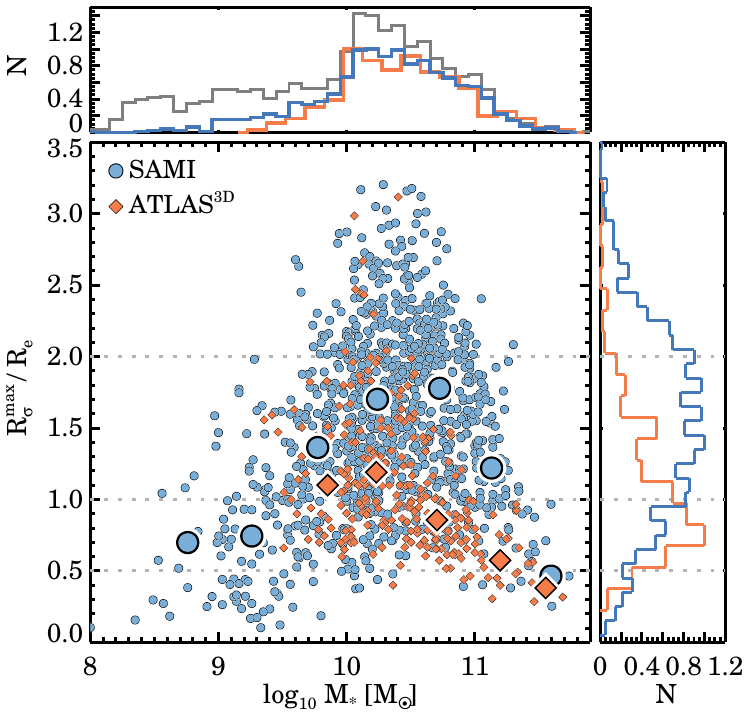}
    \caption{Comparison of the ratio between the largest aperture radius out to which the stellar kinematic data meet our quality criteria (\rmaxs) and the effective semi-major axis (\re) for galaxies in the SAMI Galaxy Survey (blue circles), and the \at\ Survey (orange diamonds). We show \rmaxs/\re\ versus stellar mass (main panel), the normalised distribution of stellar mass (top panel), and the normalised distribution of \rmaxs/\re\ (right panel). In the top panel, the full SAMI v0.9.1 stellar mass distribution is shown in grey for reference. Large symbols in the main panel show the median \rmaxs/\re\ in mass bins of 0.5 dex. This figure highlights the \emph{key problem} and main motivation for this paper: \rmaxs/\re\ depends strongly on stellar mass. At low mass ($M_* < 10^{10}$\msun), \rmaxs/\re\ is limited by spectral resolution and S/N. At high stellar mass ($M_* > 10^{11}$\msun) \rmaxs/\re\ is limited by the galaxy mass-size relation, i.e., in redshift limited surveys such as the SAMI and \at\ Survey, massive galaxies typically have larger angular sizes than the hexabundles or IFU.}
    \label{fig:rmax_stellar_mass}
\end{figure}

Fig.~\ref{fig:rmax_stellar_mass} shows the ratio of the largest aperture radius and the effective semi-major radius for galaxies in the SAMI Galaxy Survey (blue circles), and the \at\ Survey (orange diamonds). The normalised distributions in \rmaxs/\re\ are shown in the right panel Fig.~\ref{fig:rmax_stellar_mass}, which highlights the differences between the largest aperture radius in both samples. In the SAMI sample, 79 percent (767/970) have \rmaxs $>$ \re, and 23 percent (225/970) have \rmaxs $>$ 2\re, whereas for the \at\ sample 46 percent (118/259) have \rmaxs $>$ \re, and 4 percent (9/259) have \rmaxs $>$ 2\re. The distribution in stellar mass (top panel) is similar between the samples, although SAMI has a significantly larger number of galaxies at low stellar mass ($M_* < 10^{9.5}$\msun). 

Both samples, however, suffer from the same aperture bias. This is clearly visible from the triangular shaped overall distribution, and from the median \rmaxs/\re\ in stellar mass bins as indicated by the large symbols: at low and high stellar mass, \rmaxs\ quickly drops below one \re. At low stellar mass, for SAMI data, this bias is predominantly caused by low S/N, but also due to spectral resolution as \sobs\ drops below $\sim35 \kms$. At high stellar mass ($M_* > 10^{11}$\msun), both samples are limited by the size of the IFU or hexabundle, as the mass-size relation dictates that the most massive galaxies are also the largest. Thus, in order to create a sample with homogeneous stellar kinematic measurements out to one effective radius over a large range in stellar mass, it is imperative that we investigate whether aperture corrections need to be applied to those measurements.

\subsection{Extracting \vs\ and \lr\ from SAMI and \at\ Data}
\label{subsec:measurements}

\begin{figure*}
	\includegraphics[width=\linewidth]{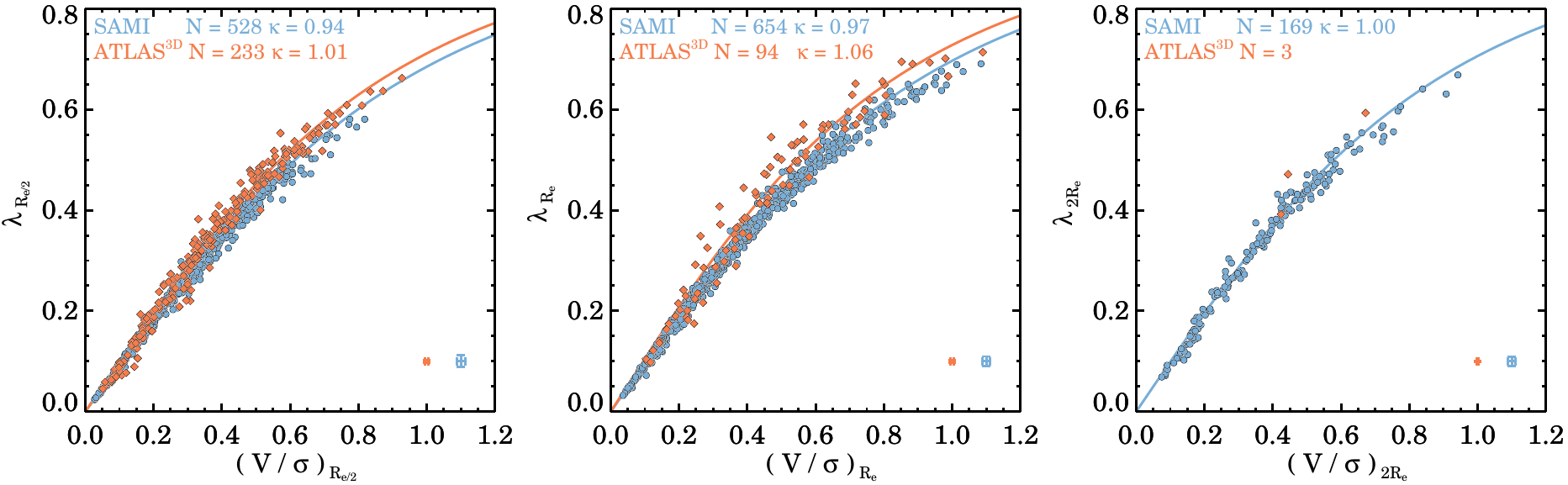}
    \caption{ \lr\ versus \vs\ in three different apertures: \ret, \re, \retwo. Galaxies from the SAMI Galaxy survey are shown as blue circles, \at\ Survey data are shown as orange diamonds. The median uncertainty is shown in the bottom-right corner. We find a tight relation between \vse\ and \lre. The different lines show the best-fitting relation between the two parameters. With increasing aperture, we find that the relation significantly steepens. The relation is steeper for \at\ data than SAMI, yet there is less scatter within the SAMI sample.}
    \label{fig:vsigma_lambdaR}
\end{figure*}

For each galaxy, we derive \vs\ using the following definition by \citet{cappellari2007}:

\begin{flushleft}
\begin{equation}
\left(\frac{V}{\sigma}\right)^2 \equiv \frac{\langle V^2 \rangle}{\langle \sigma^2 \rangle} = \frac{ \sum_{i=0}^{N_{spx}} F_{i}V_{i}^2}{ \sum_{i=0}^{N_{spx}} F_{i}\sigma_i^2},
\label{eq:vs}
\end{equation}
\end{flushleft}

\noindent and the spin parameter proxy \lr\ is derived from the following definition by \citet{emsellem2007}:

\begin{flushleft}
\begin{equation}
\lambda_{R} = \frac{\langle R |V| \rangle }{\langle R \sqrt{V^2+\sigma^2} \rangle } = \frac{ \sum_{i=0}^{N_{spx}} F_{i}R_{i}|V_{i}|}{ \sum_{i=0}^{N_{spx}} F_{i}R_{i}\sqrt{V_i^2+\sigma_i^2}}.
\label{eq:lr}
\end{equation}
\end{flushleft}

\noindent Here the subscript $i$ refers to the spaxel position within the ellipse, $F_{i}$ the flux of the $i^{th}$ spaxel, $V_{i}$ is the stellar velocity in \kms, $\sigma_{i}$ the velocity dispersion in \kms. $R_{i}$ is the semi-major axis of the ellipse on which spaxel $i$ lies, not the circular projected radius to the center as is used by e.g., \at\ \citep{emsellem2007}. We sum over all spaxels $N_{spx}$ that meet the quality cut Q1 and Q2 within an ellipse with semi-major axis \re\ and axis ratio $b/a$.

For the SAMI Galaxy Survey data, we use the unbinned flux, velocity, and velocity dispersion maps as described in Section \ref{subsec:stelkin_sami}. For the \at\ data, the unbinned flux maps are combined with the Voronoi binned stellar kinematic data as described in Section \ref{subsec:maxaperre}. We measure \vs\ and \lr\ for a large number of elliptical apertures out to \rmaxs. 

From now on we will require a stricter fill factor of 95 percent of spaxels within the aperture, as these measurements will be used for deriving aperture corrections. This restricts the analysis to smaller subsamples: N=528 at \ret, N=654 at \re, and N=169 at \retwo in the SAMI sample, whereas for \at\ we have N=233 at \ret, N=94 at \re, and N=3 at \retwo. For SAMI data we typically extract $\sim20$ different apertures, and for \at\ data $\sim50$ different apertures because of the larger number of spatial resolution elements for each galaxy. Finally, the \vs\ and \lr\ radial growth curves are interpolated at fixed apertures ranging from 0.2, 0.3, etc., out to 2.5 \re\ but never beyond \rmaxs.

In Fig.~\ref{fig:vsigma_lambdaR} we show \vs\ and \lr\ in three different apertures (\ret, \re, \retwo) for SAMI data (blue circles), and \at\ data (orange diamonds). For SAMI, we find a tight relation between \vs\ and \lr\ with little scatter, at every aperture. In the \at\ sample there appear to be more outliers, which could be due to the higher spatial resolution data in which complex dynamical features are better resolved and not washed out. We follow \citet{emsellem2007,emsellem2011} in fitting the following relation between \lr\ and \vs:

\begin{equation}
    \lr = \frac{\kappa~(\vs)}{\sqrt{1-\kappa^2~(\vs)^2}}.
	\label{eq:lr_vs}
\end{equation}

\noindent Our best-fitting relation to the SAMI data, using the $IDL$ function $MPFIT$ \citep{markwardt2009}, reveals an increasing $\kappa$ for increasing apertures: $\kappa=0.94$, $\kappa=0.97$, $\kappa=1.00$ for \ret, \re, \retwo, respectively. We find a similar trend for \ret, and \re, in the \at\ data ($\kappa=1.01$, $\kappa=1.06$, respectively), but there are too few galaxies ($N=3$) with apertures out to \retwo\  to obtain an accurate fit. The formal fitting uncertainties on $\kappa$ are small, $\sim0.001$ for SAMI data, and $\sim0.0005$ for \at, but systematic errors due to spatial resolution and seeing are not included in the fit. Our best-fitting $\kappa$ for the \at\ data out to one \re\ is lower as compared to the value given by \citet{emsellem2011}, $\kappa~\sim1.1$, which can be ascribed to our different definition of \lr\ and a different sample selection.

The best-fitting relation to the \at\ data is always higher as compared to the SAMI data. We investigate if the lower SAMI value could be caused by spatial resolution and seeing in Appendix \ref{sec:seeing_ac}, but we find that this has no significant effect on the \lr-\vs\ relation ($\Delta \kappa = -0.02$ with a $\Delta$FWHM = 0\farcs5-3\farcs0 seeing). Furthermore, for SAMI data, we find no correlation between the fit residual (data minus best-fitting model) and \re/FWHM$_{\rm{PSF}}$. However, the fit residual does correlate with S/N  and the uncertainty on \vs\ and \lr. When the uncertainties are relatively large, or when the S/N is relatively low, the offset from the best-fitting relation is more negative. \lr\ is known to be sensitive to measurement uncertainties \citep{emsellem2011,wu2014,vandesande2017}, and can be overestimated because the \lr\ calculation includes $\vert V \vert$, that can never be less than zero. As \vs\ (Eq.~\ref{eq:vs}) and \lr\ (Eq.~\ref{eq:lr}) contain $\vert V \vert$ and $V^2$ respectively, both measurements will be biased by measurement uncertainties, which is strongly correlated with S/N. However, \lr\ is normalised by $V$ and $\sigma$, whereas \vs\ is only normalised by $\sigma$. Thus, \vs\ will be more biased towards higher values than \lr\ when the S/N is low, which results in a negative offset from the \lr-\vs\ relation. The median \vs\ offset of the SAMI data from the \at\ relation is $\sim0.04$, which is still lower than uncertainties due to the impact of seeing \citep[e.g., ][]{vandesande2017}. Thus, while S/N impacts the \vs\ measurements more than \lr, it will not change the conclusions of this paper.

\begin{figure*}
	\includegraphics[width=\linewidth]{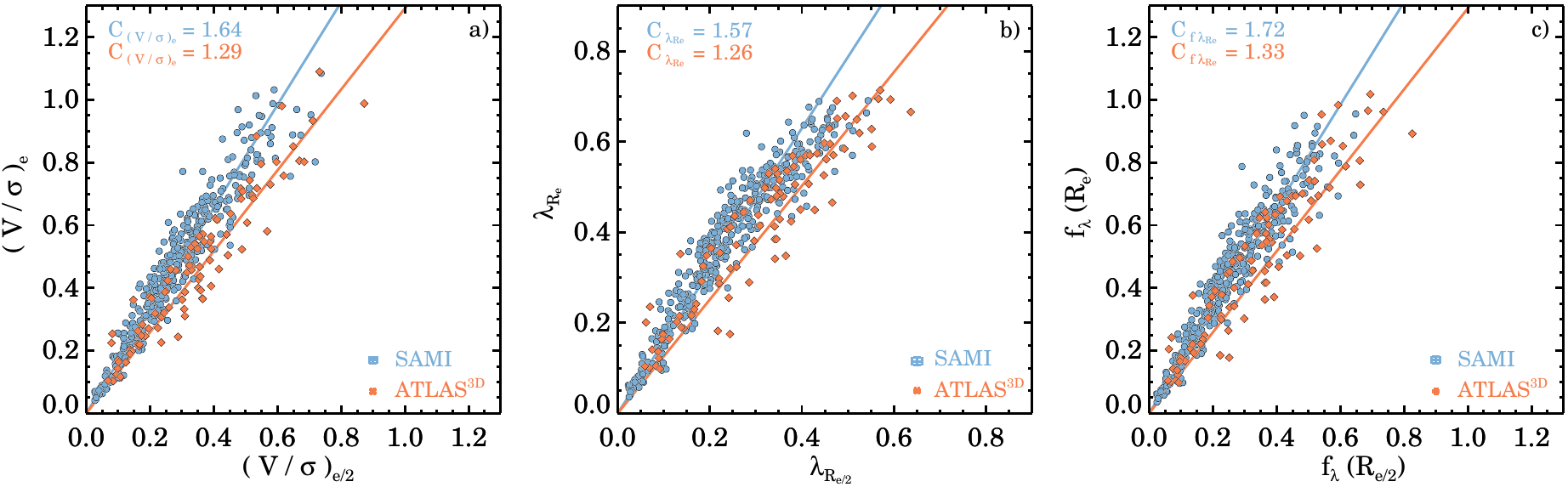}
    \caption{\vs, \lr, and \flr\ (corrected \lr) measured in different apertures \re\ and \ret. Symbols as in Fig.~\ref{fig:vsigma_lambdaR}. The median uncertainty is shown in the bottom-right corner. We find a strong linear relation between \vset\ and \vse\ in panel a), and between \flre\ and \flret\ in panel c), whereas the \lre\ and \lret\ relation is more curved. The different lines show the best-fitting relation (Eq.~\ref{eq:aperture_f_vs}, \ref{eq:aperture_lr}, and \ref{eq:aperture_f_lr}) between the two parameters, which can be used to apply a simple aperture correction. The SAMI data lie above the \at\ data, due to atmospheric seeing effects.}
    \label{fig:vs_lr_aper_comp}
\end{figure*}

We also detect a weak trend with stellar mass, such that low mass galaxies scatter more below the relation. However, as stellar mass and mean S/N are correlated in SAMI Galaxy Survey data, the trend is more likely to be caused by S/N rather than stellar mass. Another potential bias could arise from using different sources for Fig. \ref{fig:vsigma_lambdaR}a, \ref{fig:vsigma_lambdaR}b, and \ref{fig:vsigma_lambdaR}c because only 25 galaxies in the SAMI sample have reliable measurement at all of 0.5\re, 1.0\re, and 2.0\re. If we repeat the fit with only those 25 sources with full coverage, we find a small deviation ($\sim0.01$) of the best-fitting values as compared to fitting the full sample: $\kappa=0.93$, $\kappa=0.96$, $\kappa=1.02$. For \at\ there are too few sources ($N=3$) to obtain an accurate fit for all three apertures. Given that the offset is significantly smaller than the scatter in the \lr-\vs\ relation, we conclude that using different sources for different aperture comparisons does not bias our results. Similarly, by refitting a low ($z<0.05$) and high-redshift ($z>0.05$) sample, we find no significant ($\sim0.01$) deviation from the best-fitting values of the full sample.

Some of the \lr\ and \vs\ outliers were highlighted by \citet{emsellem2007,emsellem2011} to motivate that \lr\ is better than \vs\ at discriminating between fast and slow rotators. For galaxies with complex inner kinematic structures, \lr\ appeared more consistent with the overall kinematic properties than \vs. However, in the SAMI data, such outliers appear to be absent. For relatively small apertures of \ret, seeing and spatial resolution of SAMI data could wash out the impact of inner dynamical structures, but for larger apertures of \re\ and \retwo\ the tight relation between \lr\ and \vs\ persists. Furthermore, the examples used in \citet{emsellem2007,emsellem2011}, namely NGC 5813 and 3379, have limited apertures of \rmaxs/\re=0.38 and 0.50, which makes it harder to argue that at one \re, either \lre\ or \vse\ is better at classifying these galaxies as fast or slow rotators. Thus given the low number of outliers in our SAMI data, we argue that \vs\ and \lr\ have the same predictive and classification power when a consistent aperture of one \re\ is used in seeing-limited surveys. Though, as the scatter appears to be larger in the \at\ data, \lr\ could still prove to be more useful than \vs\ for classifying slow and fast rotators. However, the addition of kinemetry \citep{krajnovic2006,krajnovic2011}, Jeans anisotropic modelling \citep{cappellari2008}, radial kinematic information \citep{foster2016,bellstedt2017}, and/or high-order stellar kinematics \citep{krajnovic2011,vandesande2017}, could provide significantly more insight in the stellar kinematic properties of galaxies than using \vs\ or \lr\ alone.

\subsection{Simple Aperture Corrections}
\label{subsec:simple_ac}

Here, we explore whether a tight relation exists for \vs\ and \lr, that could be used to correct the aperture-incomplete data. We start by comparing \vs\ (Fig.~\ref{fig:vs_lr_aper_comp}a) and \lr\ (Fig.~\ref{fig:vs_lr_aper_comp}b) in two different apertures: \ret\ and \re. In the SAMI Galaxy Survey, a total of 381 galaxies simultaneously have reliable \ret\ and \re\ measurements, and 94 galaxies in the \at\ data.

 There is a tight linear correlation between \vset\ and \vse, whereas \lret\ versus \lre\ is slightly non-linear and curves downwards towards the one-to-one relation at higher \lre, most prominently visible in the SAMI data. We model the data by fitting linear relations, which are shown as the solid lines in Fig.~\ref{fig:vs_lr_aper_comp}a-b: 
\begin{equation}
    \vse=C_{\vse}~\vset,
	\label{eq:aperture_f_vs}
\end{equation}
\begin{equation}
    \lre=C_{\lre}~\lret.
	\label{eq:aperture_lr}
\end{equation}
\noindent For SAMI galaxies, the best-fitting aperture corrections are $C_{\vse} = 1.64$, $C_{\lre} = 1.57$, whereas for \at\ galaxies, the values are significantly lower: $C_{\vse} = 1.29$, $C_{\lre} = 1.26$. The vertical root-mean-square (RMS) scatter increases for larger values of \vs\ and \lr\ and is similar for \vs\ and \lr: 15.1 percent versus 15.6 for SAMI data, and 16.4 percent versus 16.8 percent for \at\ data. In Fig~\ref{fig:vs_lr_aper_comp}b we see that for $\lret>0.35$ most of the SAMI data is on the right-side of the best-fitting relation. We could use an exponential function to fit the relation between \lre\ and \lret, however, for the larger aperture \lre\ versus \lretwo, the curvature changes direction from downwards to upwards. This makes it more complicated to construct an aperture correction using one single function that describes all combinations of radii. Therefore, we remove the non-linearity in \lr\ by replacing \lr\ with \flr\ following \citet{emsellem2011}:
\begin{equation}
    \flr = \frac{\lr}{\sqrt{1-\lr^2}},
	\label{eq:flre}
\end{equation}
This equation is based on the relation between \lr\ and \vs\ (Eq~\ref{eq:lr_vs}). Next, we fit the \flre\ versus \flret\ data with the following linear relation:
\begin{equation}
    \flre=C_{\flre}~\flret.
	\label{eq:aperture_f_lr}
\end{equation}

\noindent In Fig.~\ref{fig:vs_lr_aper_comp}c, we show the relation between \flret\ and \flre\ and the best-fitting relation ($C_{\flre} = 1.72$ for SAMI, $C_{\flre} = 1.33$ for \at). The non-linearity at high values has now disappeared but the RMS scatter from the best-fitting relation is slightly higher as compared to the \lre-\lret\ relation: 16.6 percent and 19.4 percent for SAMI and \at\ data respectively. 

\begin{figure*}
	\includegraphics[width=\linewidth]{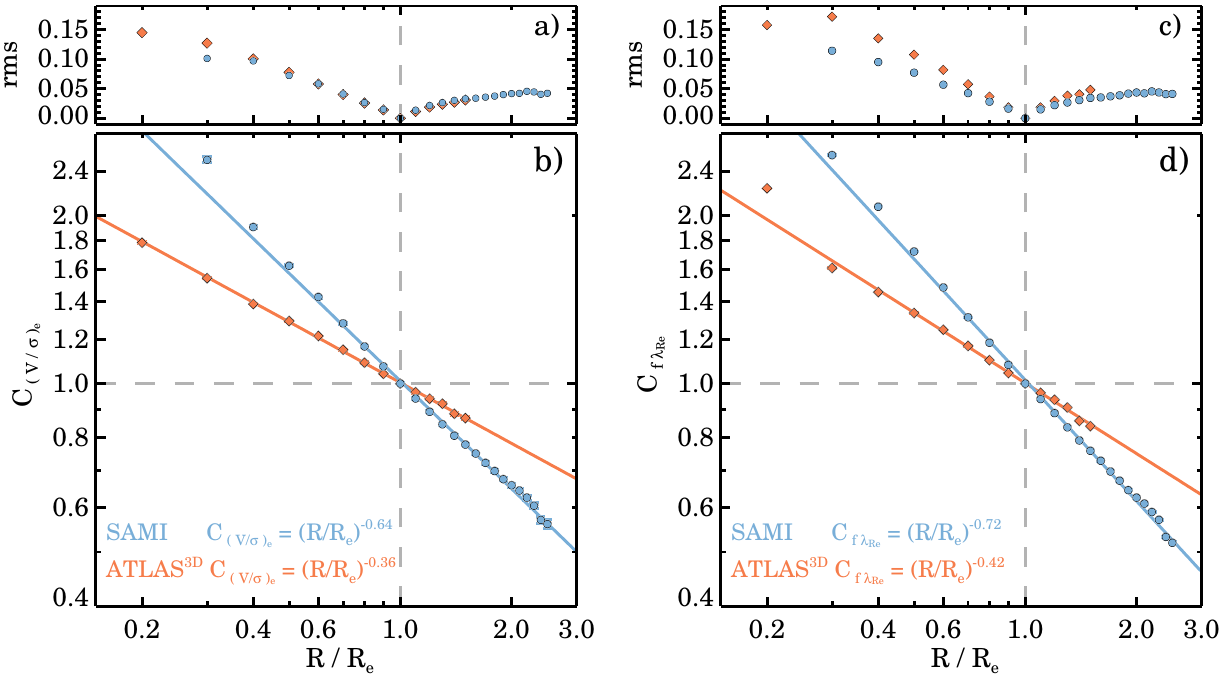}
    \caption{Aperture correction values as a function of aperture radius for \vs\ (panel b) and \lr\ (panel d). SAMI data are shown as blue circles, galaxies in the \at\ Survey are shown as orange diamonds. We show the RMS scatter of every aperture-radius fit in panel a) and c). We find a tight relation as indicated by the solid lines, that can be approximated by Eq.~\ref{eq:cvs_sami}-\ref{eq:cflr_a3d}. These relations can be used as a first-order aperture correction in seeing limited surveys (e.g., SAMI, MaNGA) and in surveys where the impact of seeing is small (e.g., \at\ \& CALIFA).}
    \label{fig:vs_lr_multi}
\end{figure*}

However, we find that our best-fitting $C$ values to the \at\ data are significantly higher than the quoted values of $C_{\vse} \sim 1.1$ and  $C_{\flre} \sim 1.15 $ in \citet{emsellem2011}. We can only recover these values for the \at\ data if all 259 \at\ galaxies are fitted, irrespective of their aperture coverage. As 57 percent of the sample have aperture radii less than one \re, consequently the relation between \lre\ and \lret\ will be artificially closer to the one-to-one relation.

In Appendix \ref{sec:seeing_ac}, we show that seeing has a significant effect on $C_{\vse}$ and $C_{\flre}$. With increasing seeing, smaller apertures (e.g., \ret) are more severely impacted as compared to larger apertures (e.g., \re) as the strongest gradients in both flux and velocity are in the centre. For seeing with FWHM = 0\farcs5 to 3\farcs0, we find an increase in $C_{\vse}$ from 1.24 to 1.50, and for $C_{\flre}$ from 1.28 to 1.50. The trend is the same as for the SAMI and \at\ data in Fig.~\ref{fig:vs_lr_aper_comp}. However, with the typical seeing for SAMI \citep[2\farcs1, ][]{allen2015}, the simulated aperture correction is lower than the observed: $C_{\vse} = 1.37$ versus 1.64 respectively, and $C_{\lre} = 1.45$ versus 1.72, respectively. The mismatch could be caused by the selected sample of galaxies used to estimate the impact of seeing, which is relatively small and has a limited range in \vs\ and \lr. A more thorough analysis of the impact of seeing on SAMI measurements is under way, but beyond the scope of this paper.

Thus, while seeing is important, by analysing both the SAMI and \at\ aperture relations, we can now work towards providing simple aperture corrections for seeing-impacted surveys (e.g., SAMI \& MaNGA) and for surveys where the impact of seeing is small (e.g., \at\ \& CALIFA). We emphasize that the key result from Fig.~\ref{fig:vs_lr_aper_comp} is that the vertical root-mean-square (RMS) scatter between both \vse-\vset\ and \flre-\flret\ is small: $\sim0.08$. Furthermore, we find no correlations between the residual of the aperture correction relation with stellar mass and effective radius of the galaxy. This suggests that it is possible to apply a simple correction to our \vs\ and \lr\ measurements when the size of the aperture is limited over the entire sample. 

Next, we fit equation \ref{eq:aperture_f_vs} and \ref{eq:aperture_f_lr} over a large range of apertures, from 0.3-2.5 $R_{\rm{aper}}/\re$ for SAMI galaxies, to 0.2-1.5 $R_{\rm{aper}}/\re$ for \at\ galaxies. We require a minimum of 10 galaxies to accurately fit the relation. Fig.~\ref{fig:vs_lr_multi} shows the best-fitting values of $C_{\vse}$ and $C_{\flre}$ as a function of aperture radius, which follows a simple tight power law. For all apertures the relation for SAMI is steeper than for \at, which we show is predominantly due to seeing (see Appendix \ref{sec:seeing_ac}). We find that the \vse\ aperture corrections can be derived from:
\begin{equation}
    C_{\vse} = (R_{\rm{aper}}/\re)^{-0.64}~~~\rm{[~SAMI~]},
	\label{eq:cvs_sami}
\end{equation}
\begin{equation}
    C_{\vse} = (R_{\rm{aper}}/\re)^{-0.36}~~~\rm{~[\at~]}.
	\label{eq:cvs_a3d}
\end{equation}
Similarly for \flre\ we find that:
\begin{equation}
    C_{\flre} = (R_{\rm{aper}}/\re)^{-0.72}~~~\rm{[~SAMI~]},
 	\label{eq:cflr_sami}
\end{equation}
\begin{equation}   
    C_{\flre} = (R_{\rm{aper}}/\re)^{-0.42}~~~\rm{[~\at~]}.
	\label{eq:cflr_a3d}
\end{equation}

\noindent In summary, from measuring the relation between different apertures for \vs\ and \flr, we derive a simple relation between the aperture correction $C_{\vse}-C_{\flre}$ and aperture radius $R_{\rm{aper}}/\re$ (Eq.~\ref{eq:cvs_sami}-\ref{eq:cflr_a3d}), that can be used to aperture correct data. We test the accuracy of this method in Section \ref{subsec:comp_methods}, and apply it to the full SAMI Galaxy Survey and \at\ Survey in Section \ref{sec:application}.

\begin{figure*}
	\includegraphics[scale=0.90]{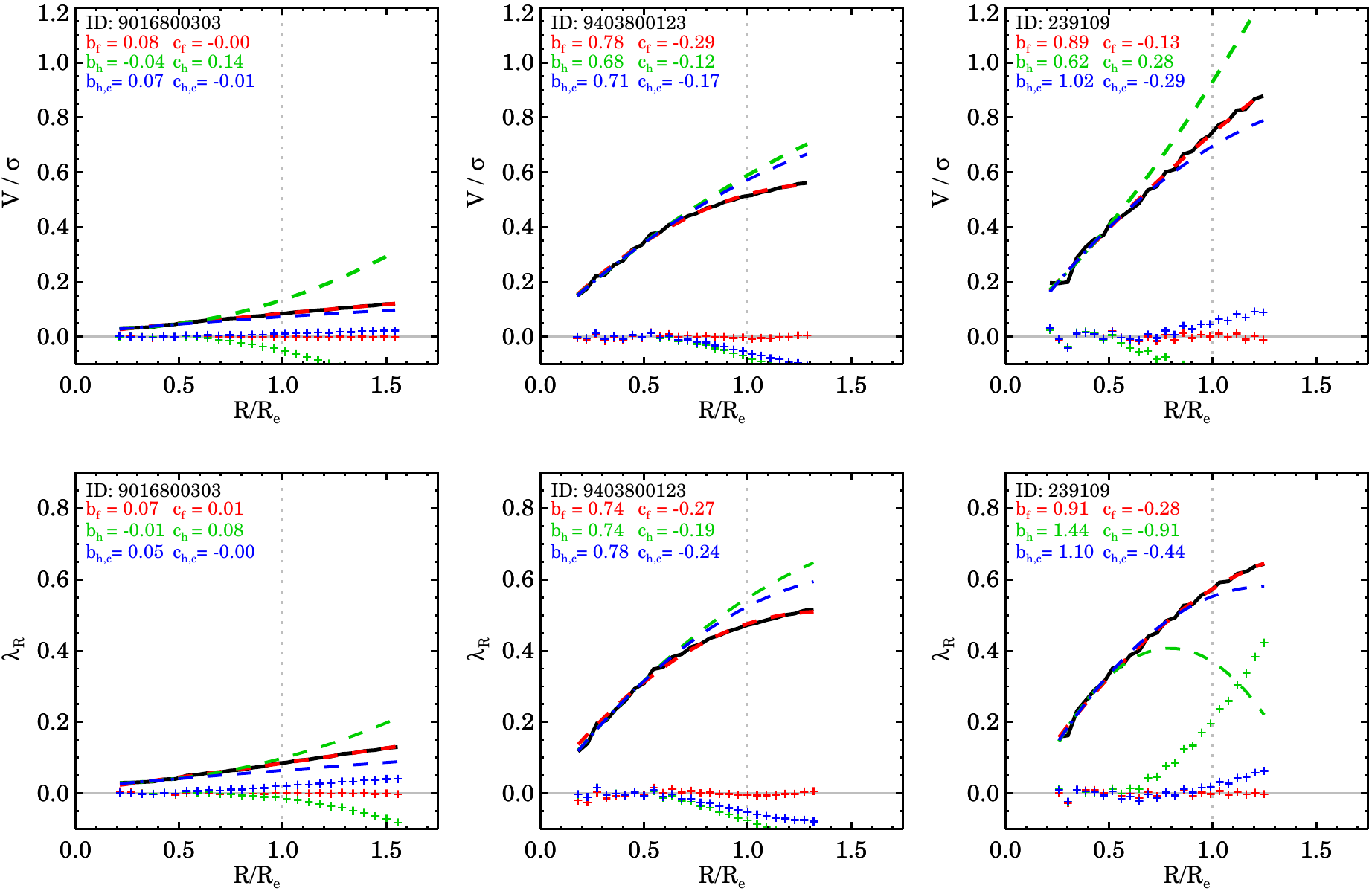}
    \caption{Growth curves of \vs\ and \lr\ in three example SAMI galaxies. We show the observed data in black, the best-fitting quadratic function to the full profile as the red dashed line, the best-fitting relation to the inner \ret\ profile in green, and in blue the best-fitting relation to the inner \ret\ profile with fitting constraints as described in Section \ref{subsec:curve_growth_ac}. Residuals (data - best fit) are shown as the plus symbols. \vs\ and \lr\ growth curves are well approximated by second order polynomials. The constrained fits out to \ret\ (blue) show that the inner profile can be extrapolated and used to recover \vs\ and \lr\ profiles out to at least one effective radius. We find that in 84 percent of the cases the fit with constraints (blue) recover the observed value at \re\ better than the fits without (green).}
    \label{fig:profile_fit}
\end{figure*}
\begin{figure*}
	\includegraphics[scale=1.35]{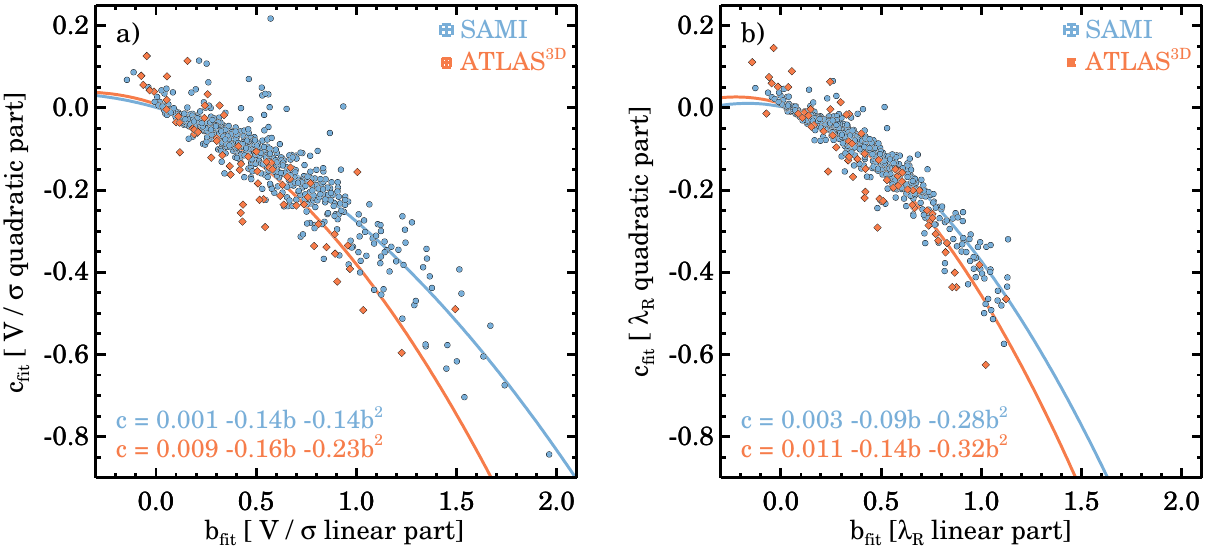}
    \caption{Best-fitting linear and quadratic parameters of all \vs\ and \lr\ profiles. Galaxies from the SAMI Galaxy survey are shown as blue circles, \at\ Survey data are shown as orange diamonds. The median uncertainty is shown in the top-right corner. There is a tight relation between $b$ and $c$ with little scatter, which suggest that the best-fitting relation between $b$ and $c$ can be used as a constraint when fitting the inner \vs\ and \lr\ growth curves.}
    \label{fig:profile_param}
\end{figure*}

\subsection{Aperture Corrections from Radial Growth Curves}
\label{subsec:curve_growth_ac}


In this section, we aim to reduce the scatter in the aperture corrections further, by extrapolating the full measured kinematic radial profile, instead of only using the largest aperture radius measurement. We start by fitting all the \vs\ and \lr\ radial growth curves with a second order polynomial:

\begin{equation}
    \vs = a + b(R/\re) + c(R/\re)^{2}
	\label{eq:vs_curve_of_growth}
\end{equation}

\begin{equation}
    \lr = a + b(R/\re) + c(R/\re)^{2}
	\label{eq:lr_curve_of_growth}
\end{equation}

In Fig.~\ref{fig:profile_fit} we show three example SAMI galaxies with their growth curves as the black solid line for \vs\ (top row) and \lr\ (bottom row), and their best-fitting second order polynomial shown in red. Residuals (data minus best fit) of the full profile fit are shown as red pluses around the zero line. For fitting we use the $IDL$ function $MPFIT$ \citep{markwardt2009}. For the SAMI data, we set the minimum radial profile aperture to contain 15 good spaxels due to seeing limitations; for \at\ we set the limit to 20 good spaxels to lower the impact of complex inner dynamics (e.g., counter rotating cores) on the growth curve fits. We find that both \vs\ and \lre\ are well-fitted by second order polynomials, with less than one percent scatter: \vs\ RMS = 0.010, \lr\ RMS = 0.008 for SAMI data ($N=629$), and \vs\ RMS = 0.009, \lr\ RMS = 0.009 for \at\ data ($N=68$).

From a visual inspection of the growth curves, it appears that galaxies with low \vs\ and \lr\ show mostly linear behaviour, whereas high \vs\ and \lr\ galaxies follow more quadratic functions. In Fig.~\ref{fig:profile_param} we investigate this further by showing the best-fitting linear parameter $b$ versus the quadratic parameter $c$ from Eq.~\ref{eq:vs_curve_of_growth}-\ref{eq:lr_curve_of_growth}. We indeed find that there is a relation between the two: if the \vs\ or \lr\ profile is slowly rising (small $b$), then the profile is mostly linear (small $c$), whereas if the profile is steeply increasing (high $b$), the profile is always more curved (lower $c$). To approximate the scatter, we fit a quadratic function between $b$ and $c$ (solid lines Fig.~\ref{fig:profile_param}), and find an RMS = 0.046 for \vs\ and RMS = 0.030 for \lr\ in the SAMI data, and an RMS = 0.074 for \vs\ and RMS = 0.058 for \lr\ in the \at\ data. 

\begin{figure*}
	\includegraphics[width=\linewidth]{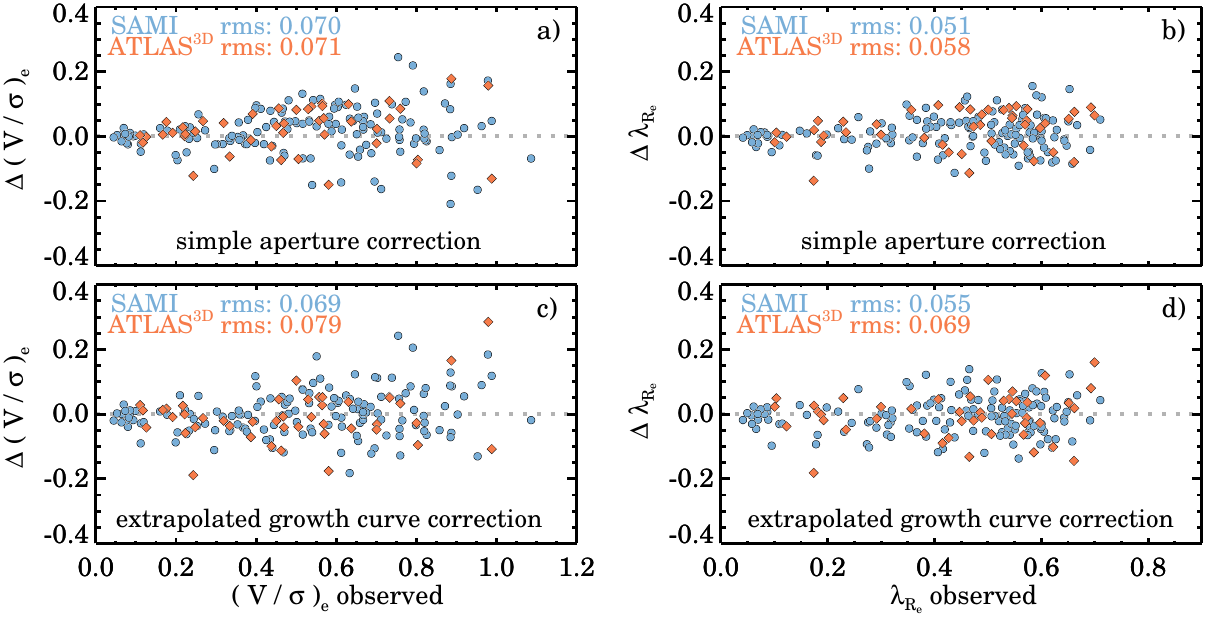}
    \caption{Comparison of the observed and aperture-corrected \vse\ and \lre\ values ($\Delta=$ observed - aperture corrected) as a function of \vse\ (left) and \lre\ (right). In the top row, we apply the simple aperture corrections as described in Section \ref{subsec:simple_ac}; the bottom rows show the growth curve method from Section \ref{subsec:curve_growth_ac}. Galaxies from the SAMI Galaxy survey are shown as blue circles, \at\ Survey data are shown as orange diamonds. From the RMS scatter we conclude that the simple aperture corrections work as well the aperture corrections from fitting and extrapolating the growth curves.}
    \label{fig:quality_fit}
\end{figure*}

The relation for \at\ galaxies lies below the one for SAMI, i.e., \at\ growth curves show a stronger quadratic behaviour as compared to the ones from SAMI. After inspecting several outliers, we find that some of the curvature is caused by a premature radial flattening of the profiles. One explanation is that this could be due to more extensive binning in the outskirts of these \at\ galaxies that could artificially lower \vs\ and \lr. When we apply a stricter aperture quality cut to the \at\ data, many of the lower outliers indeed disappear, but the sample size also decreases to $N\sim50$ which makes it harder to quantify the relation. In both the SAMI and \at\ data there are several galaxies where the profile shows a quadratic upturn ($c>0$). Nearly all of the \at\ objects with $c>0$ are classified by \citet{krajnovic2011} as non-regular rotators with kinematically-distinct cores, double maxima, or double $\sigma$ features. Thus it is unsurprising that the growth curves for these galaxies deviate from galaxies with regular rotation velocity fields.

Motivated by the tight relation between the linear and quadratic component in the \vs\ and \lr\ growth curves, we postulate that the inner profile ($<\ret$) can be used to derive a more accurate aperture correction than the single aperture correction value from Section \ref{subsec:simple_ac}. To test this theory, we first fit the inner \ret\ profiles with a second order polynomial without any constraints, shown in Fig.~\ref{fig:profile_fit} as the green dashed lines, and extrapolate beyond \ret. In the fit, we require a minimum of five radial points within \ret, which significantly lowers the number of galaxies in both samples for which we can test this method: $N=141$ for SAMI, $N=44$ for \at. 

For all three galaxies in Fig.~\ref{fig:profile_fit}, we obtain a poor match between the extrapolated profile (green) and the observed (black) beyond \ret. Therefore, to improve the fit, we now add a constraint to the parameter $c$ in Eq.~\ref{eq:vs_curve_of_growth}-\ref{eq:lr_curve_of_growth} by using the relation as given in Fig.~\ref{fig:profile_param}. Thus parameter $c$ is now coupled to parameter $b$. In other words, if the profile is slowly increasing (low value of $b$), the fit is also forced to be linear (low $c$). However, if the profile is steeply increasing (high $b$), the fit is now forced to have a more quadratic shape (higher $c$). The blue dashed line in Fig.~\ref{fig:profile_fit} shows the fit to the inner \ret\ growth curves, now including these constraints. There is a clear improvement for two galaxies as compared to fits without a constraint (green), and the extrapolated profile now more closely match the observed (black) profiles; in general we find that in 84 percent (118/141) the fit with constraints recover the observed value at \re\ better.

In summary, we show that \vs\ and \lr\ growth curves can be well approximated by a quadratic function, and that there is a tight relation between the linear and quadratic component of each profile. We use the relation between the linear and quadratic component to demonstrate that the outer profile can be extrapolated from the inner profile. This provides an alternative method for calculating aperture-corrected \vse\ and \lre\ values. In the next section we test how well this new method works as compared to the more simple aperture correction method derived in Section \ref{subsec:simple_ac}.

\subsection{Comparing Methods}
\label{subsec:comp_methods}

In the previous sections, we explored two different methods for calculating aperture-corrected \vs\ and \lr\ values. Here, we test and compare both methods on SAMI and \at\ data. The test sample is similar to the sample from the previous Section \ref{subsec:curve_growth_ac}, where we selected galaxies that have coverage out to at least one \re, with a minimum of five radial points within \ret, to extrapolate the profiles.

Our method for comparing the accuracy of the aperture corrections is as follows: first, we extract \vs\ and \lr\ within \ret\ and \re\ from the observed growth curves. Then \vset\ and \lret\ are used to calculate the aperture-corrected values at one \re\ using the method described in Section \ref{subsec:simple_ac}. The results are shown in the top row of Fig.~\ref{fig:quality_fit}, where in panel a) we compare $\Delta \vse = $ \vse\ observed - \vse\ aperture corrected, as a function of \vse\ observed, and likewise in panel b) for \lre. With increasing \vse\ and \lre\ the scatter between the observed and aperture-corrected measurements increases. This is perhaps unsurprising, as we earlier observed that the scatter in Fig.~\ref{fig:vs_lr_aper_comp} also increases for larger \vs\ and \lr. However, the absolute fractional scatter in the data is similar across \vse\ (mean 10.9 percent for SAMI data) and \lre\ values (10.5 percent for SAMI data). We find no significant difference  between the SAMI and \at\ data, which indicates that the scatter is more likely caused by the intrinsic differences in galaxies, rather than measurement  uncertainties. 

We note that by applying the wrong aperture correction to the wrong sample, e.g., Eq.~\ref{eq:cvs_a3d} on SAMI data, causes a median offset of $\Delta \vse = 0.09$. Thus applying the aperture corrections presented here to other survey data could create an artificial offset in \vs\ and \lr\ if they do not match the instrumental set-up and typical atmospheric conditions of either the SAMI or \at\ survey. For large upcoming surveys such as MaNGA, we would advise following the method outlined here to calibrate the aperture correction relations, if a subset of the data allows for multi-aperture measurements.

Next, we derive the aperture-corrected \re\ values by fitting the inner \ret\ growth curves with constraints as described in Section \ref{subsec:curve_growth_ac} (Fig.~\ref{fig:quality_fit}c-d). The trends are similar to the top-row, i.e., the scatter in the recovered values increases as a function of \vse\ and \lre. Disappointingly, we find that the scatter on average is slightly larger for the extrapolated growth curve method as compared to the simple aperture corrections. Similar results are obtained if we restrict the sample to the best-quality data, i.e., most complete spatial sampling and highest S/N; the overall RMS scatter is lower, but no significant differences are found between both methods. This suggests that our simple method of calculating aperture corrections works as well as, or even better than compared to our more complicated growth curve fitting approach. 

Overall, in Fig.~\ref{fig:quality_fit} we find that the mean fractional uncertainty on \vs\ and \lr\ at one \re\ is 11 percent when the aperture only extends out to \ret. Thus, applying an aperture correction to \vs\ and \lr\ is a significant improvement over using non-aperture-corrected data; if no aperture corrections are applied, \vs\ and \lr\ would be underestimated by a factor of 30-60 percent (Fig.~\ref{fig:vs_lr_aper_comp}).

\begin{figure*}
	\includegraphics[width=\linewidth]{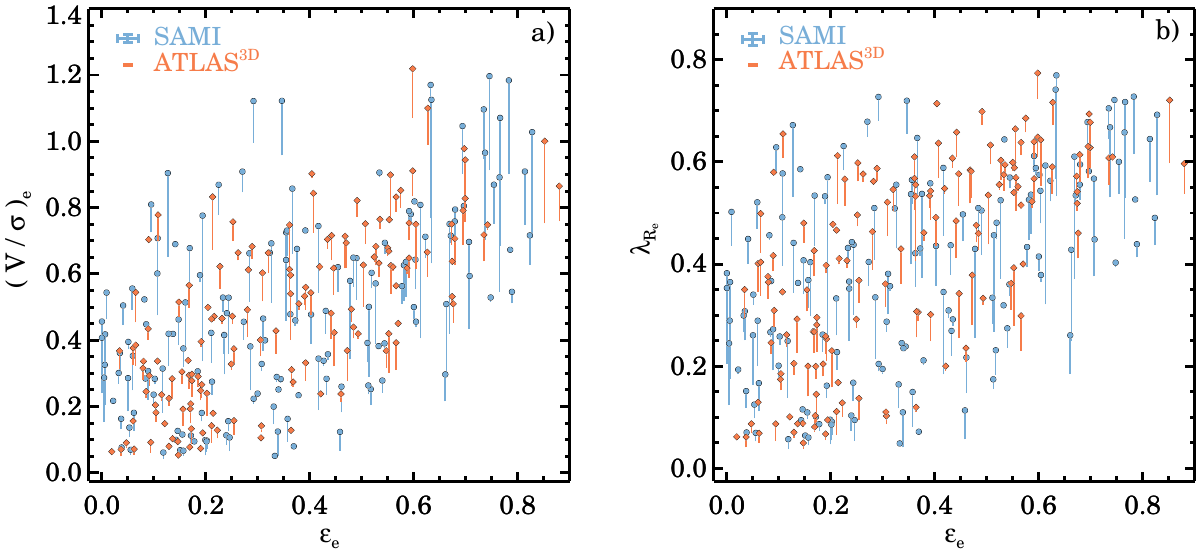}
    \caption{ \vse\ and \lre\ versus ellipticity $\epsilon_{\rm{e}}$ for all galaxies where an aperture correction is required (\rmaxs<\re). For SAMI (blue circles) and \at\ (orange diamonds) data, the aperture corrections are shown as the solid lines; the final aperture-corrected value is indicated by the filled symbols. The median uncertainty is shown in the top-left corner. We find that the aperture corrections on average significantly increase \vs\ (respectively 18, and 11 percent for SAMI and \at), and \lr\ (respectively 14, and 9 percent for SAMI and \at).}
    \label{fig:vs_lr_epsilon_lines}
\end{figure*}
\begin{figure*}
	\includegraphics[width=\linewidth]{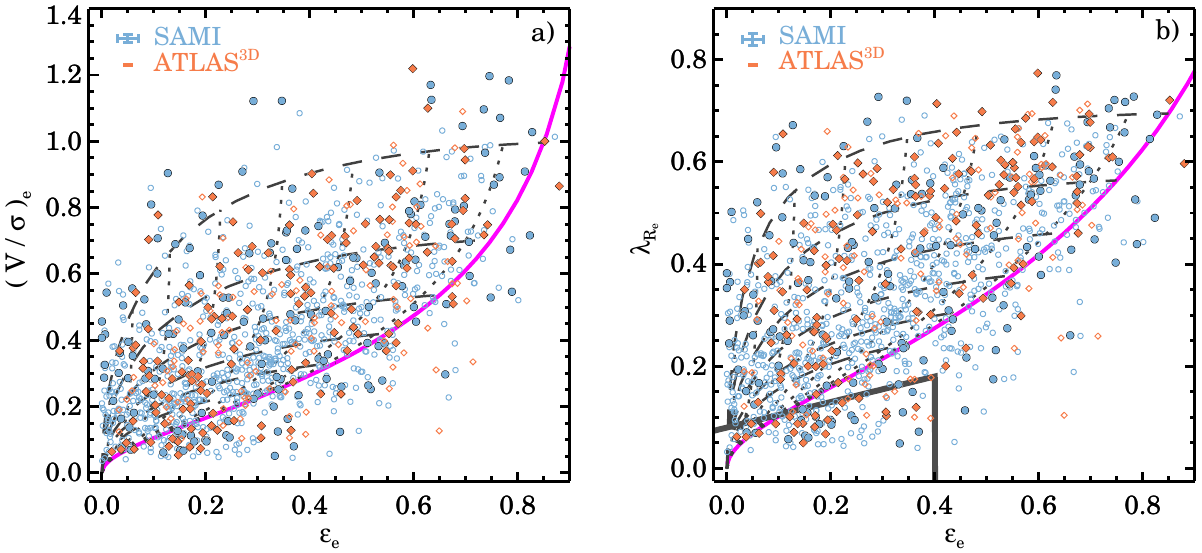}
    \caption{ \vse\ and \lre\ versus ellipticity $\epsilon_{\rm{e}}$. Aperture-corrected measurements are presented as the filled symbols, whereas the open smaller symbols show data without aperture corrections. Galaxies in the SAMI Galaxy Survey are shown as blue circles, galaxies in the \at\ Survey by orange diamonds. The median uncertainty is shown in the top-left corner. Furthermore, we show the theoretical prediction for the edge-on view of axisymmetric galaxies with $\beta_z = 0.70\times \epsilon_{\rm{intr}}$ as the solid magenta line (assuming $\kappa=0.97$). The gray dashed lines correspond to the locations of galaxies with different intrinsic ellipticities $\epsilon_{\rm{intr}}$=0.85-0.35 \citep[see][]{cappellari2007,emsellem2011}, while the dotted lines show the model with different viewing angle from edge-on (magenta line) to face-on (towards zero ellipticity). The solid grey line in panel b) shows the slow/fast rotator separation from \citet{cappellari2016}. For the sample with $\log M_{*}$/\msun $> 11$, we find that the fraction of slow rotators increases from $24.2\pm5.3$ percent to $35.9\pm4.3$ percent when aperture-corrected measurements are combined with the data without aperture corrections.}
    \label{fig:vs_lr_epsilon}
\end{figure*}
\begin{figure*}
	\includegraphics[width=\linewidth]{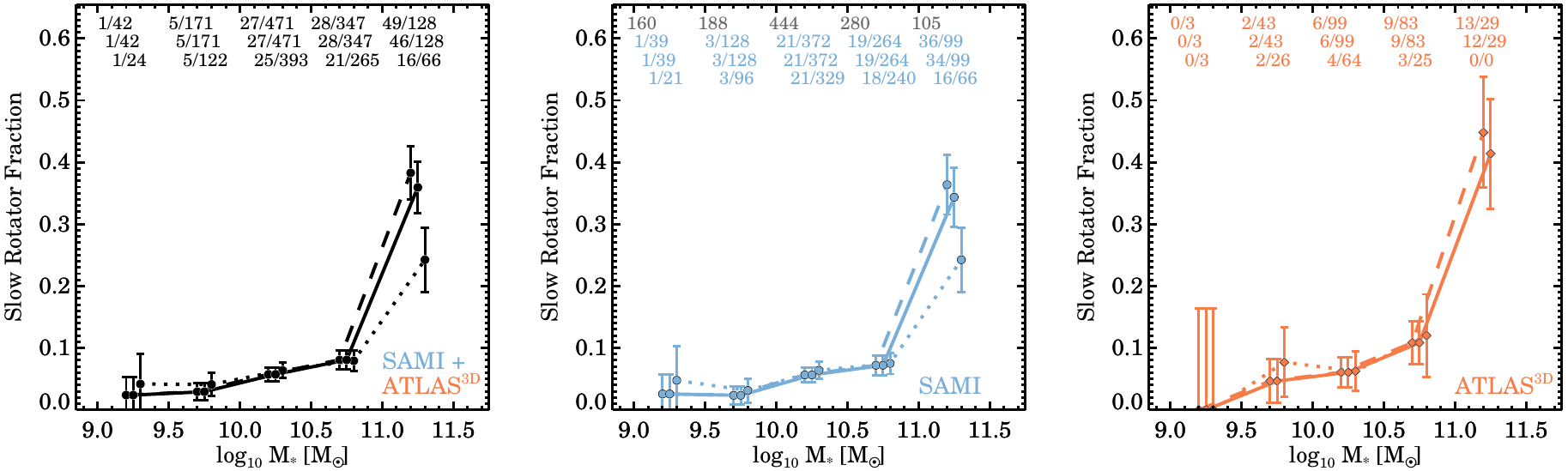}
    \caption{Fraction of galaxies classified as slow rotators versus stellar mass. We show the results for the combined SAMI Galaxy Survey and \at\ Survey in the left panel, for the SAMI Galaxy Survey (with both early- and late-type galaxies) in the middle panel, and the \at\ Survey (early-type galaxies only) in the right panel. In each panel, we show the fraction of slow rotators when measurements with \rmaxs<\re\ are not included in the sample (dotted line, numbers in bottom row), with aperture corrected applied (solid line, numbers in middle row), or when the largest aperture radius measurements are used (dashed line, numbers in top row, with the exception of the middle panel where the numbers in the top row give the total number of SAMI v0.9.1 galaxies in each mass bin). We find a strong increase in the fraction of slow rotators as a function of stellar mass. The fraction of slow rotators is underestimated if galaxies with \rmaxs<\re\ are not included in the sample, whereas with the largest aperture radius measurements we find a similar fraction of slow rotating galaxies as compared to using aperture corrected values.}
    \label{fig:sr_fraction_lmass}
\end{figure*}


\section{Application: Fraction of Slow Rotators}
\label{sec:application}

We started this paper by arguing that aperture effects are important for studying the fraction of galaxies with slow rotation as a function of stellar mass. Here, we investigate if and how the fraction of slow rotators changes if aperture-corrected data are included in this calculation. For the SAMI sample the number of galaxies increases from $N=767$ to $N=920$ when we include aperture-corrected measurements; the number of galaxies in the \at\ sample increases from $N=118$ to $N=259$. 

Before we calculate the fraction of slow rotators, in Fig.~\ref{fig:vs_lr_epsilon_lines} we first show the impact of the aperture-corrections in the \vse-\ee\ and \lre-\ee\ plane for all galaxies where the aperture coverage is insufficient. The solid lines show the total aperture correction from \rmaxs\ to \re\ as indicated by the filled symbols. For SAMI galaxies where the largest aperture radius is less than \re ($N=153$), the mean aperture correction is $\Delta \vs=0.087 $, or 18 percent, and $\Delta \lr= 0.062$ or 14 percent. For \at\ data with $\rmaxs < \re$ ($N=140$), the median aperture correction is $\Delta \vs=0.047$, or 11 percent, and $\Delta \lr= 0.036$ or 9 percent.

In Fig.~\ref{fig:vs_lr_epsilon} we show the full SAMI Galaxy Survey and \at\ Survey sample, with full \re\ measurements (open symbols) and aperture-corrected (filled symbols) for \vse\ (panel a) and \lre\ (panel b), versus the ellipticity within one effective radius \ee. We find that a large fraction of galaxies with aperture corrections populate the low \vse\ and \lre\ region, but also at high \vse\ ($> 0.6$) and \lre\ ($> 0.5$). Low \vse-\lre\ galaxies are likely large massive galaxies with little rotation, whereas the latter are big rotating disks.

We define galaxies as slow rotators by adopting the selection criteria from \citet{cappellari2016}:
\begin{equation}
\lre < 0.08+\ee/4 ~~~~~~ \rm{with} ~~~~~~ \ee<0.4.
\end{equation}

For the combined SAMI-\at\ sample, we find that the fraction of slow rotators increases from $7.8\pm1.0$ percent (69/886) to $9.2\pm0.9$ percent (108/1179) when aperture-corrected measurements are combined with the non-aperture corrected values. Confidence intervals are calculated using the method outlined in \citet{cameron2011}. If we ignore the aperture corrections and use the largest aperture radius \lr\ measurements, the fraction of slow rotators is slightly overestimated: 9.4$\pm0.9$ percent (111/1179), however this is not significantly different from using aperture corrected values. The reason for the similarity is caused by the fact that the aperture corrections are most significant for large \lr\ and \vs\ values, whereas the fast/slow rotation division is around $\lr\sim0.2$. With aperture corrections included, the fraction of slow rotators is lower in the SAMI Galaxy Survey than in the \at\ Survey: $8.6\pm1.0$ percent (79/920) versus $11.2\pm2.0$ percent (29/259), respectively. This is due to the fact that the \at\ Survey was selected to only contain early-type galaxies, whereas the SAMI Galaxy Survey sample consists of both early- and late-type galaxies and includes more low-mass galaxies.

Next, we limit the sample to massive galaxies with $\log M_{*}$/\msun $> 11$. The fraction of slow rotators increases more dramatically when aperture-corrected galaxies are included: from $24.2\pm5.3$ percent (16/66) to $35.9\pm4.3$ percent (46/128). If we ignore the aperture corrections and use the largest aperture radius \lr\ measurements for these galaxies instead, we find that the fraction of slow rotators is slightly overestimated, but not significantly: $38.3\pm4.4$ percent (49/128). With aperture corrections included, the fraction of massive slow rotators is lower in the SAMI Galaxy Survey than in the \at\ Survey: $36.4\pm4.8$ percent (36/99) versus $44.8\pm9.0$ percent (13/29), respectively.

In Fig.~\ref{fig:sr_fraction_lmass} we show the fraction of slow rotators as a function of stellar mass. The different lines show the impact of aperture effects. The dotted lines shows the fraction of slow rotators when galaxies are not included in the sample when \rmaxs\ is less than \re\ dashed lines when the largest aperture radius measurements are used, and the solid line when the aperture corrected measurements are included. Note that we offset the median mass-bin data for the three different methods to highlight the differences in the slow rotator fraction. For the highest stellar mass bin we include all galaxies with $\log M_{*}$/\msun $> 11$.

For both the SAMI Galaxy Survey (middle-panel) and \at\ Survey (right-panel) we find a strong increase in the slow-rotator fraction as a function of stellar mass, with the strongest increase at $\log M_{*}$/\msun $> 11$. The fraction of slow rotators is higher for the \at\ Survey than for SAMI Galaxy Survey data, which can be attributed to the fact that the SAMI Galaxy Survey includes both early-type and late-type galaxies. We find that the fraction is underestimated if galaxies with \rmaxs<\re\ are not included in the sample (dotted line). If largest aperture radius measurements are used (dashed line), the fraction of slow rotating galaxies is slightly higher as compared to when aperture corrected values are used (solid line), but not significantly.

Thus, these results confirm that aperture corrections are important when calculating the fraction of slow rotating galaxies as a function of stellar mass, but that selection effects (i.e., excluding galaxies with $\rmaxs<\re$) have a significantly stronger impact on the fraction than aperture corrections. In order to assess how much the fraction could furthermore change to due selection effects, in the middle panel of Fig.~\ref{fig:sr_fraction_lmass}, we provide the total number of galaxies in the SAMI v0.9.1 sample (grey numbers, top row), as compared to the total number of galaxies with (aperture corrected) stellar kinematic measurements (blue numbers, middle row). In the most massive bin, we are nearly complete with a success rate of 99 galaxies with stellar kinematic measurements out of 105 galaxies in the parent sample. At lower stellar mass, the incompleteness increases as we no longer reach the S/N requirements to accurately measure the LOSVD parameters. Above a stellar mass of $\log M_{*}$/\msun $> 10.5$, however, the fraction of slow rotators is not significantly going to change due to selection effects.

\section{Conclusions}
\label{sec:conclusions}

In this paper, we present two methods for aperture correcting 2D stellar kinematic \vs\ and \lr\ measurements using integral field spectroscopic data from the SAMI Galaxy Survey and \at\ Survey. The necessity for aperture-correcting data is demonstrated by showing that there is a strong bias in the largest kinematic aperture radius as a function of stellar mass (Fig.~\ref{fig:rmax_stellar_mass}), and from the fact that \vs\ and \lr\ increase rapidly out to at least an effective radius (\re). 

We measure \vs\ and \lr\ for a large number of apertures in the SAMI and \at\ data, and show that there is a tight relation for both $V/\sigma$ and $\lambda_{R}$ between different apertures (Fig.~\ref{fig:vs_lr_aper_comp}). The coefficient of the relation between different \vs\ and \lre\ apertures follows a simple power law (Fig.~\ref{fig:profile_fit}), that can be used as first-order aperture correction (Eq.~\ref{eq:cflr_sami}-\ref{eq:cflr_a3d}).
 
Spatial resolution and seeing have a strong impact on the amplitude of the aperture correction (Appendix \ref{sec:seeing_ac}). In worsening seeing, the relation between small and large apertures becomes steeper as the inner profile is more strongly affected by the point spread function than the outskirts. However, because we calculate aperture corrections for both SAMI and \at\ data separately, this work provides aperture corrections for all seeing-impacted surveys where the typical seeing is $\sim2^{\prime\prime}$ (e.g., SAMI \& MaNGA) and for surveys where the impact of seeing is small (e.g., \at\ \& CALIFA).

We explore a second method for providing more accurate aperture correction based on fitting \vs\ and \lr\ growth curves of individual galaxies. \vs\ and \lr\ radial growth curves are well approximated by second-order polynomials out to 1.5\re, with little scatter (RMS $<1$ percent). We show that we can we successfully recover the profile out to one \re, from fitting the inner profile (0.5\re), but only if a constraint between the linear and quadratic parameter is applied. 

Using data with full \re\ coverage, we demonstrate that if the aperture only extends out to \ret, the simple aperture correction method and the radial growth curves can both recover \vs\ and \lr\ at one \re\ with a mean uncertainty of 11 percent. However, our simple first-order approach for calculating aperture corrections works slightly better than the more complicated approach of fitting and extrapolating the inner profile. The methods presented here provide a significant improvement over using non-aperture-corrected data, as the mean ratio between \ret\ and \re\ is a factor of 1.3-1.6 for \vs\ and \lr, which is significantly larger than the mean uncertainty of the aperture corrections.

We investigate how the fraction of fast and slow rotating galaxies changes as a function of stellar mass with and without aperture-corrected data. For the SAMI Galaxy Survey and ATLAS$^{\rm{3D}}$ survey, the fraction of slow versus fast rotating galaxies with $\log M_{*}$/\msun $> 11$ changes from $24.2\pm5.3$ percent (16/66) to $35.9\pm4.3$ percent (46/128) when aperture data are included. However, by using measurements out to the largest aperture radius we find a slow rotator fraction of $38.3\pm4.4$ percent (49/128), similar as compared to using aperture corrected values. Thus, our works suggest that when the IFS observations do not have coverage out to one \re, it is better to use largest aperture radius measurements of \vs\ and \lr, rather than excluding such galaxies from the sample, if a mass complete sample is required. As recent studies show that mass is the main driver of the kinematic morphology-density relation in clusters \citep{veale2017b,brough2017}, and with cosmological simulations that are beginning to explore the evolution of spin as a function of redshift \citep{naab2014,choi2017,penoyre2017}, this emphasizes the need for using spatially homogeneous, or aperture-corrected measurements when investigating these trends.

\section*{Acknowledgements}

We thank the anonymous referee for the very constructive comments which improved the quality of the paper. The SAMI Galaxy Survey is based on observations made at the Anglo-Australian Telescope. The Sydney-AAO Multi-object Integral-field spectrograph (SAMI) was developed jointly by the University of Sydney and the Australian Astronomical Observatory, and funded by ARC grants FF0776384 (Bland-Hawthorn) and LE130100198. JvdS is funded under Bland-Hawthorn's ARC Laureate Fellowship (FL140100278). SB acknowledges the funding support from the Australian Research Council through a Future Fellowship (FT140101166). M.S.O. acknowledges the funding support from the Australian Research Council through a Future Fellowship (FT140100255). Support for AMM is provided by NASA through Hubble Fellowship grant \#HST-HF2-51377 awarded by the Space Telescope Science Institute, which is operated by the Association of Universities for Research in Astronomy, Inc., for NASA, under contract NAS5-26555.

The SAMI input catalog is based on data taken from the Sloan Digital Sky Survey, the GAMA Survey and the VST ATLAS Survey. The SAMI Galaxy Survey is funded by the Australian Research Council Centre of Excellence for All-sky Astrophysics (CAASTRO), through project number CE110001020, and other participating institutions. The SAMI Galaxy Survey website is http://sami-survey.org/.

GAMA is a joint European-Australasian project based around a spectroscopic campaign using the Anglo-Australian Telescope. The GAMA input catalogue is based on data taken from the Sloan Digital Sky Survey and the UKIRT Infrared Deep Sky Survey. Complementary imaging of the GAMA regions is being obtained by a number of independent survey programs including GALEX MIS, VST KiDS, VISTA VIKING, WISE, Herschel-ATLAS, GMRT and ASKAP providing UV to radio coverage. GAMA is funded by the STFC (UK), the ARC (Australia), the AAO, and the participating institutions. The GAMA website is: http://www.gama-survey.org/.

Based on data products (VST/ATLAS) from observations made with ESO Telescopes at the La Silla Paranal Observatory under program ID 177.A-3011(A,B,C). Funding for SDSS-III has been provided by the Alfred P. Sloan Foundation, the Participating Institutions, the National Science Foundation, and the U.S. Department of Energy Office of Science. The SDSS-III web site is http://www.sdss3.org/.



\bibliographystyle{mnras}
\bibliography{jvds_sami_aperture_corr}



\appendix
\section{Effect of Seeing on Aperture Corrections}
\label{sec:seeing_ac}
%
\begin{figure*}
	\includegraphics[width=\linewidth]{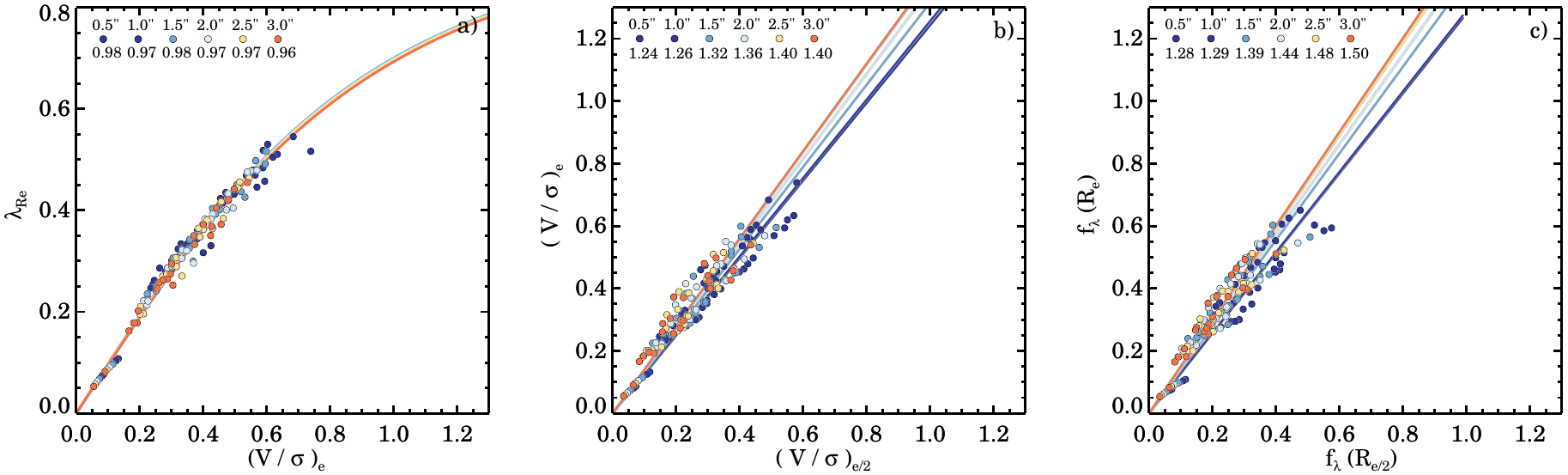}
    \caption{Relation between \vse-\lre (a),  \vset-\vse\ (b), and \flret-\flre\ (c) for 23 galaxies from the \at\ "re-observed" with SAMI under different simulated seeing conditions. Different realizations of the seeing are shown by different colours, from 0\farcs5 in blue to 3\farcs0 in red. The typical seeing for the SAMI Galaxy Survey is 2\farcs1. The different values are the best-fitting relation between  between the two parameters. The difference between \vset\ and \vse, and \lret\ and \lre\ is larger with increasing seeing values.}
    \label{fig:lr_aper_sims}
\end{figure*}

In Fig.~\ref{fig:vsigma_lambdaR} and \ref{fig:vs_lr_aper_comp}, we found that SAMI and  \at\ data show different trends, most likely due to the impact of spatial resolution and seeing. Here, we follow the same approach as outlined in \citet{vandesande2017} where we use existing \at\ kinematic measurements to study the effect of seeing and measurement uncertainty on SAMI observations. Only galaxies that have full coverage out to at least one effective radius are included. We furthermore only use galaxies where the binned data has been derived from four or less original spaxels, in order to avoid step functions in the velocity and dispersion maps. A total of 23 galaxies meet these selection criteria, which have a broad range in \lr\ (0.05-0.6) and ellipticity (0.05-0.6) \citep{emsellem2011}. 

The details of creating SAMI mock observations are described in \citet{vandesande2017}. In short, we rebin the flux, velocity, and velocity dispersion maps to get similar angular size distribution as SAMI galaxies. The effect of seeing is mimicked by constructing three dimensional flux weighted LOSVD cubes, that are convolved with a Gaussian with FWHM ranging from 0.5$^{\prime\prime}$, 1.0$^{\prime\prime}$, ..., 3.0$^{\prime\prime}$.
For each simulated galaxy, we measure \vs\ and \lr\ in different apertures as described in Section \ref{subsec:measurements}. Fig.~\ref{fig:lr_aper_sims}a-c shows the results for \vs\ and \lr\ under different simulated seeing conditions. Different colours show different realizations of the seeing, from 0\farcs1 in blue to 3\farcs0 in red. We note that typical seeing for the SAMI Galaxy Survey is $\sim2^{\prime\prime}$, indicated by the beige data.

We do not find a strong impact of seeing on the relation between \lre\ and \vse\ (Fig.~\ref{fig:lr_aper_sims}a). With increasing seeing (FWHM=0\farcs5-3\farcs0) the relation becomes less steep ($\kappa= 0.98-0.96$). However, the difference due to seeing ($\Delta \kappa = 0.02$) is significantly less than the difference we find between the SAMI and \at\ data ($\Delta \kappa = 0.09$). The effect of seeing is much stronger when we compare \ret\ and \re\ aperture measurements. For both \vs\ and \lr\ the relation becomes steeper with increasing seeing, as the inner \ret\ profile is more affected than the outer profile.


\bsp	
\label{lastpage}
\end{document}